# What increases (social) media attention: Research impact, author prominence or title attractiveness?

Olga Zagovora[*], Katrin Weller[**], Milan Janosov[***], Claudia Wagner[****] and Isabella Peters[*****]

[*] *olga.zagovora@gesis.org* ; [**] *katrin.weller@gesis.org;*
GESIS – Leibniz Institute for the Social Sciences, Cologne (Germany)

[***] *janosov_milan@phd.ceu.edu*
Central European University, Budapest (Hungary)

[****] *claudia.wagner@gesis.org;*
GESIS – Leibniz Institute for the Social Sciences, Cologne & University of Koblenz-Landau, Koblenz (Germany)

[*****] *ipe@informatik.uni-kiel.de*
ZBW – Leibniz Information Centre for Economics & Kiel University, Kiel (Germany)

**Introduction**
The most widely accepted ways of measuring the impact of academic publications are based on citations, although, recently, altmetrics (Priem et al., 2012) have been proposed as supplements to the traditional approach. One of the challenges in altmetrics is to better understand why some publications receive notable attention in, e.g., news outlets, social media platforms or other alternative sources for measuring scholarly impact – and others not. In this paper we operationalize "media attention" as mentions of publications in news outlets as well as on Twitter, Facebook, Wikipedia, scientific blogs, YouTube, Google+, StackOverflow and reddit.

It has been shown that media attention is not always directly related to traditional citations (Haustein et al., 2014). Moreover, coverage of scientific publications varies across altmetric aggregators selected for data collection and analysis (e.g. Altmetric.com, PlumX; Jobmann et al., 2014) but also across disciplines and social media-platforms (Twitter, Facebook, Mendeley, Figshare, Google+, and Wikipedia). Therefore, in-depth analysis of publication features has to consider the particular discipline or media-specific particularities.

To explore new potential explanations for media attention we develop a model that describes factors which may have an impact on the attention a paper receives. We differentiate between three categories of paper features which we will refer to as "research impact", "author prominence", and "title attractiveness".

*Research impact.* Papers that are mentioned by traditional media (news outlets) or by other users on social media platforms may be considered of high scientific importance, i.e. one may expect that these publications could receive many citations. On the one hand, Thelwall & Nevill (2018) revealed that Scopus citation counts are correlated with Mendeley readers in all fields, and Altmetric.com scores are related to citation counts in some fields. On the other hand, a meta-analysis (Erdt et al., 2016) across more than 40 cross-metric validation studies showed a weak correlation between altmetrics and citation counts; i.e. only some publications with high citation counts are more likely to appear on Twitter or in news articles. Thus, these contradictive statements, i.e., findings by Thelwall & Nevill (2018) and Erdt et al. (2016), have to be studied in more detail. Moreover, there are only few studies (Manisha & Mahesh, 2015) that are concerned with the effect of news articles on the citation counts.

*Author prominence.* Well-known authors usually receive many citations due to their reputation in their field of study (Tahamtan et al., 2016). Moreover, a relationship between the international and national cooperation of authors and the frequency of citations has been observed (Chinchilla-Rodriguez et al., 2016). Quantity and quality of collaborators is important: for example, (1) co-authorship with domestic authors may decrease number of citations (Goldfinch et al., 2003) and (2) correlation direction may vary for different fields (Didegah & Thelwall, 2013). Since these phenomena have been observed for traditional metrics, the degree of collaboration and centrality of authors in the collaboration network might be also a confounding factor for altmetrics. Thus, we complement our model with network centrality measures to analyse collaboration patters.

*Title attractiveness.* Zahedi & Haustein (2018) showed that Mendeley readership counts have a small negative correlation with the title length of paper in most disciplines. Publication titles with punctuation marks (e.g. colons) correlate with the number of citations (Jacques & Sebire, 2010). Subotic & Mukherjee (2014) found that the title amusement level and shorter titles were associated with more citations and article downloads. In our study, we will model sentiment analysis features (polarity and subjectivity), title length and usage of a colon in the title as attractiveness features to confront this finding with the results from other media sources and an Altmetric.com score.

**The objective of our study** is to reveal which features of publications are associated with increased media attention. Hence, we tackle the following research questions for two major multidisciplinary journals.:
(1) Which features do papers that receive high media attention have in common?
(2) Which discipline-specific differences exist between those features that are related with high media attention?
(3) Are the same types of publications popular across different media channels?

**Data & Method**
Next we explain the data sets and methods used in the study of the relationship between certain features of scientific papers and the media attention these papers receive.

*Datasets*
We collected 59,804 papers from the Web of Science (WoS) database[1]: 43,921 papers published from 2004 to 2017 in the Proceedings of the National Academy of Sciences (PNAS), and 15,883 papers published from 2010 to 2017 in Nature Communications (NC). Both journals are ranked among the top five best multidisciplinary journals based on their total cites and journal impact factors[2]. We used high impact multidisciplinary journals to (1) study disciplinary differences and to (2) control for paper quality; we assume that these journals only accept high-quality research (Tahamtan et al., 2016). Each journal was studied separately to control for journal-specific practices (e.g., number of co-authors, frequencies of collaborations, and title length). We studied all years since creation of NC (i.e., 2010-2017) and only papers published later than 2004 in PNAS assuming that earlier research would not be picked up that extensively on the Internet (Costas et al., 2015). The publication year was also utilised as a control variable in our multiple regression models. Thus, the data analysis was not corrupted by year-particularities which are substantial in bibliometrics and altmetrics (Haustein et al., 2014).

---
[1] https://www.webofknowledge.com (accessed 07.12.2017)

The WoS data contains the following information for every paper: publication year, title, full names of authors, citation counts and DOIs. This information was utilised for further data collection to build a complex feature set describing each paper. The research topic was assigned to each publication: for PNAS via the publisher website[3]; for NC via Springer Nature SciGraph Data Explorer[4]. Then, each publication was labelled with exactly one OECD field of science[5] which outcome is represented in Table 1.

For each paper Altmetric.com[6] was used to obtain (a) the number of mentions in mainstream media outlets, Twitter, Wikipedia and Facebook public posts and (b) the Altmetric.com score, a weighted approximation of the general attention across different social media and on-line media outlets. The Altmetric.com scores were transformed to the Normalised Log-transformed Altmetric Score (NLAS) with $\ln(1 + x_i)/l_i$, where $x_i$ is the Altmetric.com score and $l_i$ is the arithmetic mean of the log-transformed $\ln(1 + x)$ Altmetric.com scores from the same discipline and year as $x_i$. This transformation, analogously to the Mean Normalized Log-transformed Citation Score (MNLCS) (Thelwall & Fairclough, 2017), reduces skewing and eliminates year and discipline-specific practices of media promotion.

Table 1. Number of publications used in study

| Field of science (OECD) | Number of papers, PNAS | Number of papers, NC |
|---|---|---|
| Agricultural sciences | 221 | 62 |
| Engineering & Technology | 1133 | 2482 |
| Humanities | 0 | 35 |
| Medical & Health sciences | 12813 | 2649 |
| Natural sciences | 27994 | 10410 |
| Social sciences | 1810 | 245 |
| **Total** | **43921** | **15883** |

*Feature selection*
We differentiate between three types of features that could potentially influence the media attention a publication receives. The first group of features reflects the impact of the papers assuming the quality of the research (research impact), the second group of features is based on the authors' attributes (author prominence), and the third group of features is derived from the linguistic properties of the publication titles (title attractiveness). In the following we will outline our approach for quantifying the three dimensions.

---

[2] Thomson Reuters Journal Citation Reports 2016 https://jcr.incites.thomsonreuters.com/ (accessed 01.12.2017)
[3] http://www.pnas.org/ (accessed 04.08.2017)
[4] https://scigraph.springernature.com (accessed 07.12.2017)
[5] https://www.oecd.org/science/inno/38235147.pdf (accessed 01.03.2018)
[6] https://www.altmetric.com/ (accessed 07.12.2017)

*Research impact* is reflected by the number of citations for both journals obtained from WoS in 12/2017. To account for year- and discipline-specific citation practices, MNLCS was utilised. Year of publication was utilised as a control parameter, too, since more recent publications would receive more social media mentions and less citations than older publications (Costas et al., 2015).

*Author prominence* looked at authors with specific roles, namely first and last authors. We adopt the common assumption that for many publications, unless the alphabetical order was used, the person with the most contribution is the first author and the last authorship is given to a principal investigator. To quantify author prominence, we calculated the PageRank for first and the last authors based on the collaboration network. PageRank is a numerical weighting to each element of a network, with the purpose of quantifying its relative importance within the dataset (Page et al., 1999). The collaboration network is a graph that represents the collaboration between authors. Two authors collaborated if they were both listed as authors of the same publication. A PageRank of an author $A$ in this network is defined by the PageRanks of those authors who wrote papers together with $A$. PageRank contribution of an author $B$ to the author $A$ is always weighted by the number of co-authors of the author $B$ and so on. This means that the more collaborators the author $B$ has, the higher benefit author $A$ will receive.

*Title attractiveness* was approximated by the following title characteristics: (a) length of a title (word counts), (b) usage of a colon to separate the title from the subtitle, (c) subjectivity of the title, and (d) polarity (sentiment) of the title. However, we do not aim to judge whether, for example, a title with or without colon is more attractive, but rather take this feature as one potential characteristic that publications with high media attention might have in common.

The subjectivity of a sentence is a weighting between 0 and 1. A completely subjective sentence would have the weight of 1 and objective would be 0. For subjective titles sentiment is evaluated using polarity score. The polarity score ranges between -1 and 1, and identifies whether the expressed opinion in a title is positive (values close to 1), negative (values around -1), or neutral (approximately 0). Subjectivity and polarity were calculated using Python library TextBlob[7].

*Modelling*
We utilised multiple linear regression models to identify the features of papers with increased media attention. The following parameters were used as dependent variables: NLAS, number of publications mentioned in mainstream media outlets, tweet and retweet counts, number of Facebook public posts, Wikipedia articles counts which mention particular scientific outcome. Since multicollinearity was detected between some parameters, the final list of independent variables contains: MNLCS, year of publication, title length, usage of colon in the title, title subjectivity, title polarity, PageRank of the first and the last author.

---

[7] http://textblob.readthedocs.io/en/dev/advanced_usage.html (accessed 12.04.2018)

The disciplinary differences were studied in separate multiple linear regression models. The same independent variables as described above were utilised. NLAS was used as a dependent variable. We excluded publications in Humanities due to few data points in both journals.

**Results**
The results of multiple linear regression models among the different media sources attention scores are represented in Table 2. The correlation coefficients show that publications with higher citation impact (MNLCS), shorter and more subjective titles are more likely to be covered in social media as measured by NLAS. We also found a positive correlation between positive sentiment and NLAS, and posts counts in news outlets as well as Twitter for NC journal. In other words, NC journal papers with a positive sentiment in the title are more likely to be referenced in the mass media news articles and in tweets than papers with a negative sentiment. The correlation coefficients can be interpreted as follows: a publication in the NC journal, in which the sentiment polarity of the title equals 1, will have NLAS of 0.2154 units higher on average than a publication in NC with the polarity -1, given that the polarity of the title is the only difference of the two papers from NC journal. A publication from NC with 10 words in the title will be referenced in approximately 1.8 more news outlets on average than a publication with 20 words in the title, given that all other publication parameters are the same and the papers are published in NC journal.

Publications from PNAS are more likely to receive higher media attention if the title is composed of parts separated by a colon. At the same time, NC publications have a higher probability of being referenced on Wikipedia if the title is composed of parts separated by a colon.

If the last author of PNAS paper has a low PageRank, their publications are more likely to (1) receive high media attention, (2) be mentioned in some news article, (3) be mentioned in Facebook public post (about NC publication too) and (4) be tweeted more frequently (about NC publication too). One might interpret PageRank results as follows: (last) authors who published once in a small team or on their own are more likely to receive high media attention than those who collaborated with many different teams. One explanation could be that the publication, which was interesting for (general) public and thus received high media attention, was co-authored by a person that is external to established scientific community. In contrary, we have not found statistically significant correlations between the first author PageRank and NLAS or media posts counts for NC publications. Thus, one cannot generalise that a person who collaborates less receives high media attention; rather we could argue that new collaborators, who are assigned as the last author in the paper and are new researchers to the community, attract media attention to their research.

Correlation coefficients revealed that newer publications have a higher media attention, more news publications, tweets and Facebook posts than older publications. Only Wikipedia tends to reference older publications. These findings confirm the previous results of Kousha & Thelwall (2017).

Moreover, a publication (from NC or PNAS), given that all the paper features are zero (including zero citation impact), is more likely to be referenced on Wikipedia than on the other sources, according to Intercept correlation coefficients.

Table 2. Linear Regression model. Dependent variable: NLAS and posts counts from different media sources

|  | NLAS | News | Twitter | Wikipedia | Facebook |
|---|---|---|---|---|---|
| **Nature Communications** | | | | | |
| MNLCS | **0.1733*** | **4.1531*** | **15.4401*** | **0.1052*** | **1.3289*** |
| Length of title | **-0.0147*** | **-0.1776*** | **-0.6694*** | **-0.0035*** | **-0.0328*** |
| Title polarity | **0.1077*** | **1.5735*** | **10.62**** | 0.0253 | 0.5640 |
| Title subjectivity | **0.0427**** | 0.7098 | 0.4828 | 0.0159 | -0.0655 |
| Title chunks (":") | -0.0682 | 0.4066 | 6.4794 | **0.3613**** | 0.2058 |
| PageRank First Author | -322.8541 | 6331 | -6.554 | -2 | -3268 |
| PageRank Last Author | 726.9596 | -1747 | **-14710*** | 39 | **-12770**** |
| Year of publication | **0.0124*** | **1.3194*** | **4.8481*** | **-0.0194*** | **0.1328*** |
| Intercept | **-24.2787*** | **-2655.6*** | -9756.0 | **39.1*** | **-267.0*** |
| R$^2$ | 0.037 | 0.042 | 0.036 | 0.010 | 0.009 |
| **PNAS** | | | | | |
| MNLCS | **0.8697*** | **2.7648*** | **20.9804*** | **0.6137*** | **2.0459*** |
| Length of title | **-0.0155*** | **-0.0690*** | **-0.5533*** | **-0.0140*** | **-0.0332*** |
| Title polarity | 0.0317 | 0.0541 | 1.1126 | -0.0608 | -0.0489 |
| Title subjectivity | **0.0518*** | **0.4625*** | **3.7418*** | 0.0545 | **0.2317**** |
| Title chunks (":") | **0.0689*** | **0.7268*** | **6.0570*** | 0.0063 | **0.3226**** |
| PageRank First Author | 566 | 2653 | -1541 | 1140 | 536 |
| PageRank Last Author | **-916*** | **-15430*** | **-10280*** | 815 | **-8751*** |
| Year of publication | **0.0179**** | **0.3661*** | **1.8185*** | **-0.0266*** | **0.0144*** |
| Intercept | **-35.96*** | **-736.4*** | **-3660.7*** | **53.33*** | **-261.84*** |
| R$^2$ | 0.087 | 0.068 | 0.038 | 0.004 | 0.021 |

*p < 0.05; **p < 0.01; ***p < 0.001

Table 3. Linear Regression model grouped by OECD field of science. Dependent variable: NLAS

|  | Agricult | Eng&Tech | Med&Health | Natural | Social |
|---|---|---|---|---|---|
| **Nature Communications** | | | | | |
| MNLCS | **0.4171**** | **0.0948**** | **0.2493*** | **0.1651*** | 0.1079 |
| Length of title | **-0.0408**** | **-0.0113*** | **-0.0155*** | **-0.0107*** | -0.0055 |
| Title polarity | -0.1261 | 0.0145 | 0.0947 | **0.1119*** | 0.1335 |
| Title subjectivity | 0.0931 | 0.0004 | 0.0491 | 0.0167 | 0.0820 |
| Title chunks (":") | 0.0000 | -0.0366 | -0.1340 | 0.0392 | 0.0000 |
| PageRank First Author | 853 | -525 | -1406 | 144 | -4059 |
| PageRank Last Author | 1236 | 909 | 816 | 313 | 1546 |
| Year of publication | 0.0149 | **0.0195**** | **0.0232*** | **0.0318*** | 0.0132 |
| Intercept | -28.74 | **-38.53*** | **-45.92*** | **-63.35*** | -25.74 |
| R$^2$ | 0.146 | 0.011 | 0.022 | 0.020 | 0.012 |
| **PNAS** | | | | | |
| MNLCS | **0.6486**** | **0.6412*** | **0.9531*** | **0.8640*** | **0.6394*** |
| Length of title | -0.0172 | **-0.0121*** | **-0.0174*** | **-0.0132*** | -0.0040 |
| Title polarity | 0.4539 | 0.0073 | 0.0724 | 0.0323 | -0.1038 |
| Title subjectivity | -0.0038 | 0.1009 | 0.0259 | **0.0523**** | 0.0706 |
| Title chunks (":") | 0.0161 | -0.0633 | 0.0439 | **0.0721*** | -0.0044 |
| PageRank First Author | -4648 | 2306 | **1900*** | -11 | 1096 |
| PageRank Last Author | -3705 | 284 | **-917*** | -551 | **-2879*** |
| Year of publication | **0.0354**** | **0.0251*** | **0.0155*** | **0.0164*** | **0.0325*** |
| Intercept | **-70.91**** | **-50.34*** | **-31.32*** | **-32.99*** | **-65.09*** |
| R$^2$ | 0.118 | 0.074 | 0.100 | 0.078 | 0.203 |

*p < 0.05; **p < 0.01; ***p < 0.001

The disciplinary differences are presented in Table 3. One can see that NC publications in Agriculture have almost four times stronger positive effect of citation impact on the NLAS than publications in Engineering or Natural Sciences. In other words, a publication with the average citation impact, i.e., MNLCS equals 1, in Agricultural Sciences will receive 0.32 (0.4171 – 0.09848 ≈ 0.32) units higher NLAS than a publication with the average citation impact in Engineering & Technology, given that all other paper features are the same and the papers are published in NC journal.

However, distributions of MNLCS and row citation counts for Agricultural and Natural Sciences look very similar (with 0.87 mean MNLCS and approximately 27 citation counts), mean NLAS is higher for Agricultural Sciences (0.7 vs. 0.8). Moreover, 98% of articles in Agricultural science have non-zero Altmetric Attention Score, whereas only 84% in Natural Sciences. This evidence reveals that publications in Agriculture not only have stronger relations with citation impact but also receive more media attention than, for example, Natural Science papers.

PageRank correlation coefficients reveal interesting insights about collaborations along disciplines and media popularity. Papers in Medical and Health Sciences from PNAS are more likely to receive high media attention if the first author of a publication collaborated a lot with different teams which also publish their results in PNAS or the last author is a "newcomer" who has not yet collaborated with many diverse research teams.

Papers in Natural sciences from NC journal that have positive sentiment in the publication titles are more likely to receive media attention than publications with negative or neutral titles; from PNAS publications that have subjective titles or titles with colons have high NLAS, i.e. high media attention.

**Discussion & Conclusion**
This paper provided insights on features on research impact, title attractiveness, author prominence and their relation to media attention. We compared these features among different scientific fields, media sources, and two popular multidisciplinary journals. Thus, this paper extended ideas from previous research (Erdt et al., 2016) by adding collaboration network features and linguistic properties of titles.

We found that publications with positive sentiment in the titles are more likely to be tweeted or retweeted; moreover, publications with positive titles published in the NC journal have a higher probability of receiving media attention than publications with negative titles. This phenomenon was also observed in non-scientific tweets; findings of Ferrara & Yang (2015) suggest that Twitter users are more inclined to share and favorite tweets with positive content. Our other results, with regard to title attractiveness, are in line with findings from (Zahedi & Haustein, 2018; Jacques & Sebire, 2010) who showed that publications that use colon in the title and have shorter titles are more likely to be downloaded from publisher website, to appear in (social) media and to receive citations. In addition, the PNAS publications in which a new person appears as co-author for the first time and holds the last authorship position attract more media attention than publications of scientists who publish frequently in PNAS and collaborate with diverse co-author teams.

*Limitations.* First, the relationship of media attention and publication features may strongly be influenced by journal marketing strategies. Thus, studies should control for the journal name. Second, labelling publications with only one OECD field of science may have biased the results towards interdiciplinary publications, considering that some papers should have rather been classified to several disciplines which might have effects on MNLCS and NLAS values. Third, the PageRank does not account for different publication behaviour of disciplines. However, disciplines were studied in separate models and thus disciplinary differences with regard to PageRank were revealed, our results might be still affected in the regression models with media posts counts (e.g., Twitter) as dependent variables. Here, PageRank scores should be field normalised or the field of science has to be added as a control variable to the models.

*Future work.* Further features will be included in the analysis, such as authors' academic age and productivity level or titles' characteristics (e.g., declarative, descriptive, or interrogative title) to reveal whether 1) researchers who have more publications are more likely to have larger personal networks and higher social media attention and that was already shown for citations (Tahamtan et al., 2016) and 2) an informative title increases the popularity of a paper in the media.

**Acknowledgements**
This work was supported by the DFG-funded research project *metrics (project number: 314727790).

**References**
Chinchilla-Rodriguez, Z., Miguel, S., Perianes-Rodriguez, A., Ovalle-Perandones, M. A., & Olmeda-Gomez, C. (2016). Autonomy vs. dependency of scientific collaboration in scientific performance. *21st International Conference on Science and Technology Indicators - STI 2016*. https://doi.org/10.4995/STI2016.2016.4543

Costas, R., Zahedi, Z., & Wouters, P. (2015). Do altmetrics correlate with citations? Extensive comparison of altmetric indicators with citations from a multidisciplinary perspective. *Journal of the Association for Information Science and Technology*, *66*(10), 2003–2019. https://doi.org/10.1002/asi.23309

Didegah Fereshteh, & Thelwall Mike. (2013). Determinants of research citation impact in nanoscience and nanotechnology. *Journal of the American Society for Information Science and Technology*, *64*(5), 1055–1064. https://doi.org/10.1002/asi.22806

Erdt, M., Nagarajan, A., Sin, S.-C. J., & Theng, Y.-L. (2016). Altmetrics: an analysis of the state-of-the-art in measuring research impact on social media. *Scientometrics*, *109*(2), 1117–1166. https://doi.org/10.1007/s11192-016-2077-0

Ferrara, E., & Yang, Z. (2015). Quantifying the effect of sentiment on information diffusion in social media. *PeerJ Computer Science*, *1*, e26. https://doi.org/10.7717/peerj-cs.26

Goldfinch, S., Dale, T., & DeRouen, K. (2003). Science from the periphery: Collaboration, networks and "Periphery Effects" in the citation of New Zealand Crown Research Institutes articles, 1995-2000. *Scientometrics*, *57*(3), 321–337. https://doi.org/10.1023/A:1025048516769

Haustein, S., Peters, I., Sugimoto, C. R., Thelwall, M., & Larivière, V. (2014). Tweeting biomedicine: An analysis of tweets and citations in the biomedical literature. *Journal of the Association for Information Science and Technology*, *65*(4), 656–669. https://doi.org/10.1002/asi.23101

Jacques, T. S., & Sebire, N. J. (2010). The impact of article titles on citation hits: an analysis of general and specialist medical journals. *JRSM Short Reports*, *1*(1), 2. https://doi.org/10.1258/shorts.2009.100020

Jobmann, A., Hoffmann, C. P., Künne, S., Peters, I., Schmitz, J., & Wollnik-Korn, G. (2014). Altmetrics for large, multidisciplinary research groups: Comparison of current tools. *Bibliometrie - Praxis Und Forschung*, *3*(1), 1–19.

Kousha, K., & Thelwall, M. (2017). Are wikipedia citations important evidence of the impact of scholarly articles and books? *Journal of the Association for Information Science and Technology*, *68*(3), 762–779. https://doi.org/10.1002/asi.23694

Manisha, M., & Mahesh, G. (2015). Citation pattern of newsworthy research articles. *Journal of Scientometric Research*, *4*(1), 42–45. https://doi.org/10.4103/2320-0057.156022

Page, L., Brin, S., Motwani, R., & Winograd, T. (1999). *The PageRank Citation Ranking: Bringing Order to the Web* (Technical Report No. 1999–66). Stanford InfoLab. Retrieved from http://ilpubs.stanford.edu:8090/422/

Priem, J., Piwowar, H. A., & Hemminger, B. M. (2012). Altmetrics in the wild: Using social media to explore scholarly impact. *ArXiv:1203.4745 [Cs]*. Retrieved from http://arxiv.org/abs/1203.4745

Subotic, S., & Mukherjee, B. (2014). Short and amusing: The relationship between title characteristics, downloads, and citations in psychology articles. *Journal of Information Science*, *40*(1), 115–124. https://doi.org/10.1177/0165551513511393

Thelwall, M., & Fairclough, R. (2017). The Accuracy of Confidence Intervals for Field Normalised Indicators. *Journal of Informetrics*, *11*(2), 530–540. https://doi.org/10.1016/j.joi.2017.03.004

Thelwall, M., & Nevill, T. (2018). Could scientists use Altmetric.com scores to predict longer term citation counts? *Journal of Informetrics*, *12*(1), 237–248. https://doi.org/10.1016/j.joi.2018.01.008

Zahedi, Z., & Haustein, S. (2018). On the relationships between bibliographic characteristics of scientific documents and citation and Mendeley readership counts: A large-scale analysis of Web of Science publications. *Journal of Informetrics*, *12*(1), 191–202. https://doi.org/10.1016/j.joi.2017.12.005